\documentclass[12pt]{amsart}
\usepackage{epic,eepic}
\usepackage{pb-diagram,epsf}
\newlength{\baseunit}               % the basic unit length
\theoremstyle{definition}

\theoremstyle{remark}

\begin{document}
\title{Computing Conformal Structure of Surfaces}  

\author{Xianfeng~gu${\,}^\dag{\,}$ and  Shing-Tung~Yau ${\,}^\ddag{\,}$  }
\thanks{${\,}^\dag{\,}$Division of Engineering and Applied Science, Harvard University, Cambridge, MA. Email: gu@eecs.harvard.edu }
\thanks{${\,}^\ddag{\,}$Mathematics Department, Harvard University, Cambridge, MA. Email : yau@math.harvard.edu }
\thanks{This is a simplified version of the paper. The full version is available upon requests.}
\thanks{This research project is solely supported by Geometric Informatics Inc.}
\maketitle

\begin{center}
\begin{figure}[h!]
\begin{tabular}{ccc}
\epsfxsize=0.30\linewidth\leavevmode\epsfbox{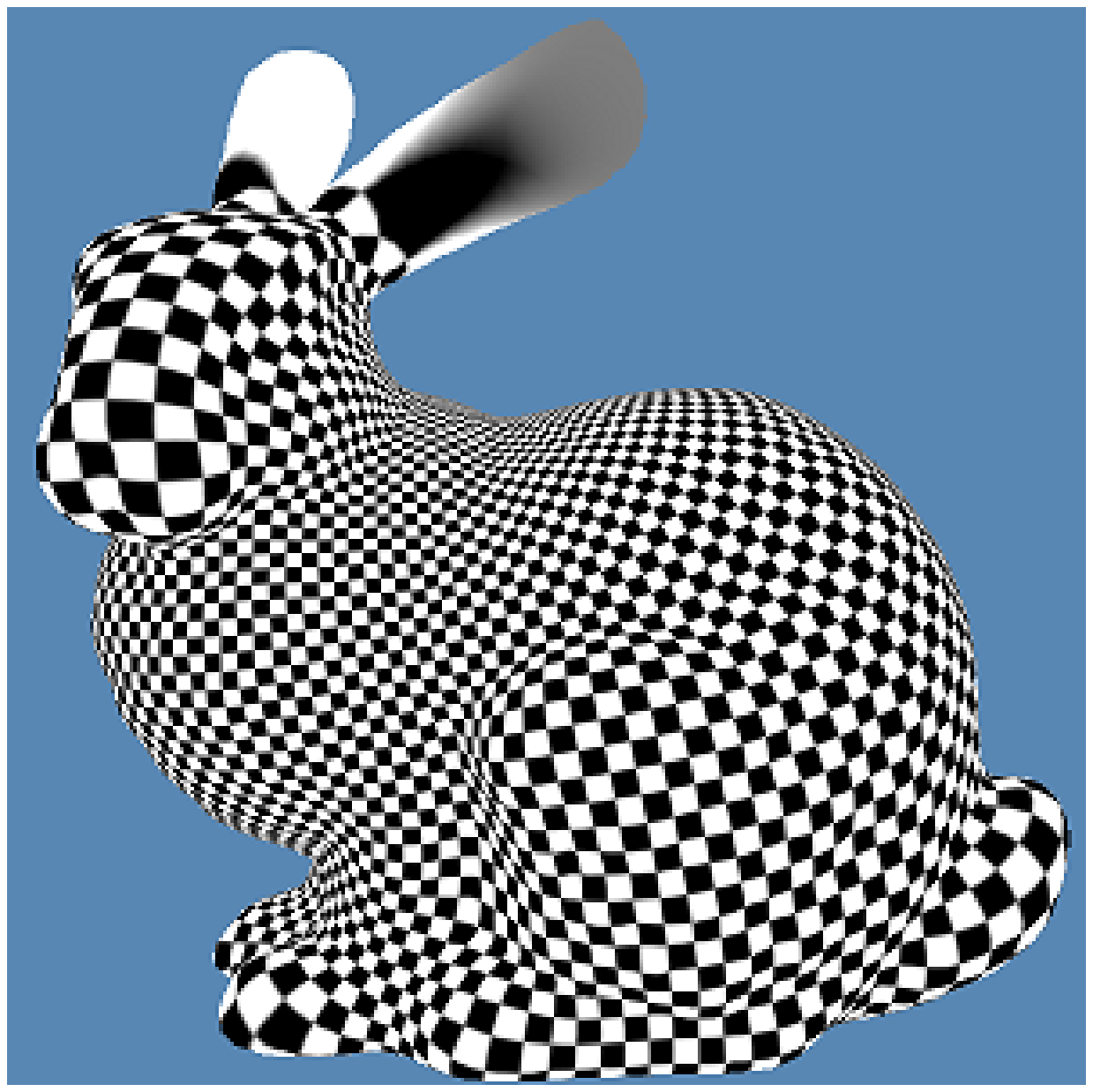}&
\epsfxsize=0.30\linewidth\leavevmode\epsfbox{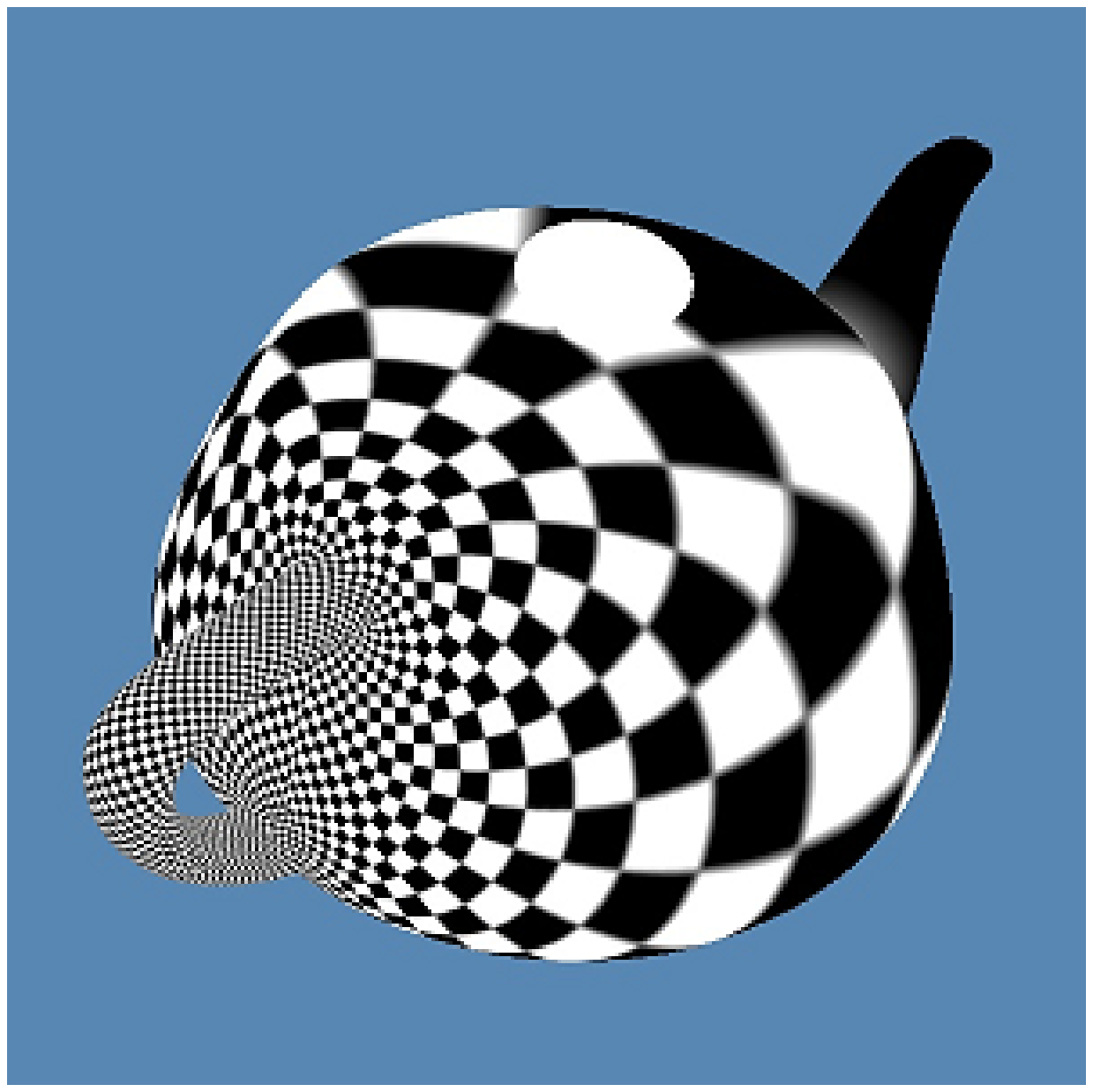} &
\epsfxsize=0.30\linewidth\leavevmode\epsfbox{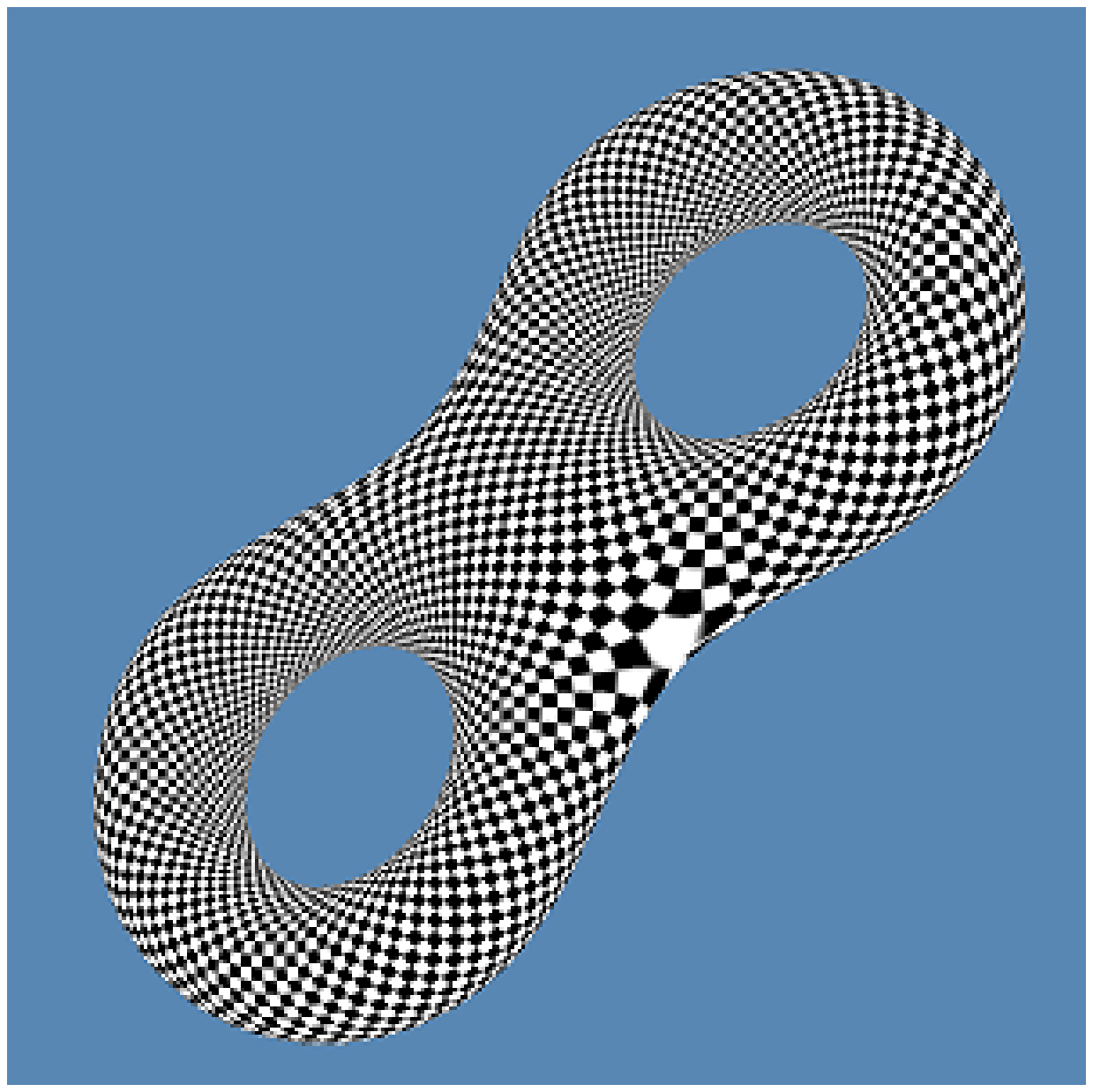}\\
(a). Genus 0 & (b). Genus 1& (c). Genus 2
\end{tabular}
\end{figure}
\end{center}

\begin{abstract}
This paper solves the problem of computing conformal structures of general 2-manifolds represented as triangle meshes. We compute conformal structures in the following way: first compute homology bases from simplicial complex structures, then construct dual cohomology bases and diffuse them to harmonic 1-forms. Next, we construct bases of holomorphic differentials. We then obtain period matrices by integrating holomorphic differentials along homology bases. We also study the global conformal mapping between genus zero surfaces and spheres, and between general meshes and planes. Our method of computing conformal structures can be applied to tackle fundamental problems in computer aid design and computer graphics, such as geometry classification and identification, and surface global parametrization.
\end{abstract}

\section{Introduction}

% \PARstart{A}{demo} file is ....
% \PARstart{A}{}demo file is ....
% \PARstart{T}{his demo} file is ....
This paper introduces a systematic way to compute conformal structures of general 2D surfaces, including computing holomorphic differentials, period matrices and conformal maps among surfaces. To the best of our knowledge, this is the first paper to give a set of practical algorithms to compute conformal structures for general closed meshes. This method has the potential to be generalized to work on meshes with boundaries and other representations of surfaces, such as implicit surfaces and level sets.

Computational conformal geometry is an active field in mathematical research. The following objects are equivalent to one another:
\begin{enumerate}
\item Compact Riemann surfaces;
\item Projective algebraic curves;
\item Jacobian varieties of compact Riemann surfaces.
\end{enumerate}
Our goal is to computationally convert these different representations from one to another, and to compute conformal invariants and conformal mappings among surfaces.

In ~\cite{seppala:curve}, ~\cite{seppala:period} and ~\cite{seppala:surface}, Riemann surfaces are represented as algebraic curves or $D/G$, where $D$ is the hyperbolic space and $G$ is a Fuchsian group acting in $D$. The homology bases are constructed as Mobius transformations, then the holomorphic differentials are found by using algebraic geometry techniques on algebraic curves. Finally the period matrices are computed explicitly by integrating holomorphic differentials on homology bases.

In real applications, geometric surfaces are represented as meshes. It is rare to represent general surfaces as algebraic curves or quotient spaces. With the development of 3D data acquisition system, large scale meshes for real objects are becoming more and more common. The above methods for abstract representation of Riemann surfaces can not be applied directly. In this paper, we assume the input data are general meshes and propose a systematic way to compute their conformal structures.

In ~\cite{stephenson} an algorithm is introduced to use circle packing to approximate conformal mappings between planar regions. For general surfaces, circle packing only considers topological structures but not geometric structures. It can not find the conformal mapping from a surface to the plane. However, if the triangulation is equilateral for each face, then the circle packing result is conformal. But, in general, such triangulation is difficult to construct. Therefore, circle packing method is not appropriate for our purpose.

In computer graphics, surface parametrization has been studied by many researchers. Hoppe et al ~\cite{Hoppe:multi} use local harmonic maps for surface simplification and editing. Haker et al ~\cite{Haker:sphere} develop an algorithm to conformally map a genus zero surface to a sphere by solving a linear system. In Haker et al ~\cite{Haker:sphere}, the sphere is stereo-graphically projected to the complex plane implicitly. The stereo projection is nonlinear in nature, large errors are introduced in the neighborhood of the north pole by using piecewise linear mapping to approximate it in practice.

In \cite{Desbrun}, Desbrun et al use conformal mapping to define geometry maps, where they compute the conformal maps from a topological disk to the complex plane. An equivalent algorithm is developed by Maillot et al ~\cite{Maillot}, who use conformal mapping for the purpose of non-distorted texture-mapping. Their method is based on Riemann-Cauchy equation. So far, although conformal mappings of genus zero surfaces have been studied, no one has tried to compute global conformal mapping for non zero genus surfaces.

In this paper, we solve the problem of computing  conformal structure of surfaces thoroughly. For genus zero surfaces, we introduce a new method to construct conformal mappings from them to spheres directly. This method avoids the stereo projection and is more stable and more accurate. More importantly, this method can be generalized to compute conformal mappings between any arbitrary two genus zero surfaces. For surfaces with non zero genus, the computation is much more complicated. We give a set of general algorithms to compute their conformal structures, which include ways to compute holomorphic differentials, period matrices and conformal mappings.

A map between two surfaces is a conformal map if it just scales the first fundamental form and preserves angles everywhere. If there is an invertible conformal map between two surfaces and the inverse is also conformal, then we call these two surfaces conformally equivalent. The conformal automorphisms form a group. The invariants under conformal transformation group are called conformal structure. Our goal is to compute these conformal structures. In terms of surface classification, conformally equivalent classes are finer than topologically equivalent classes and coarser than isometric classes.

Geometric objects classification and identification have been studied for years. But it is still an open problem today. It is challenging to classify general surfaces efficiently. A good algorithm should satisfy the following requirements: The method is intrisincly dependent of geometry and independent of triangulation; The method is stable in the sense that perturbation of geometry perturbs the result continuously; The method should also be robust enough to tolerate different resolution and boundaries; For database indexing, each class index should be small for storage and easy to compute.

Conformal mapping has many nice properties and it is suitable for classification  problems. Conformal mapping only depends on the Riemann metric and is independent of triangulation. Conformal mapping is continuously dependent of Riemann metric, so it works well for different resolutions. Conformal invariants can be represented as a complex matrix, which can be easily stored and compared. We propose to use conformal structures to classify non-zero genus surfaces. For each conformally equivalent class, we can define canonical parametrization for the purpose of comparison.

Geometry matching can be formulated to find an isometry between 2 surfaces. By computing conformal parametrization, the isometry can be obtained easily. For surfaces with close metric, conformal parametrization can also give the best geometric matching result.

\subsection{Preliminaries}

In this section, we give a brief summary of concepts and notations.

Let $K$ be a simplicial complex whose topological realization $|K|$ is homeomorphic to a compact 2-dimensional manifold. Suppose there is a piecewise linear embedding
\begin{equation}
F:|K| \to R^3.
\end{equation}
The pair $(K,F)$ is called a triangular mesh and we denote it as $M$. The q-cells of $K$ are denoted as $[v_0,v_1,\cdots,v_q]$. 

Because $M$ has a simplicial complex structure, we can compute the simplicial homology $H_*(K,R)$ and cohomology $H^*(K,R)$. We denote the {\it chain complex} as $C_*K = \{ C_qK,\partial_q \}_{q \geq 0 }$, and {\it cochain complex} as $C^*K = \{C^qK,\delta^q\}_{q \geq 0}$, where $C^qK = Hom( C_qK;R)$, the coboundary operator satisfies
\begin{equation}
\delta^q \omega \sigma = \omega \partial_{q+1} \sigma,
\end{equation}
where $\omega \in C^qK$ and $\sigma \in C_{q+1}K$.
The kernel of $\partial_q$ is $Z_qK$, the image of $\partial_{q+1}$ is $B_q$, and the {\it q-th homology group} is
\begin{equation}
H_qK = Z_q K/ B_qK.
\end{equation}
Similarly, the kernel of $\delta^q$ is $Z^qK$, the image of $\delta^{q-1}$ is $B^qK$, and the {\it q-th cohomology group is}
\begin{equation}
H^q = Z^q K/B^qK.
\end{equation}
The embedding $F$ endows $M$ with a differential structure. We then define the local charts of $M$ as $(T_i,\phi_i)$,
\begin{equation}
\label{E:chart}
\phi_i : T_i \to R^2,
\end{equation}
where $T_i$ is a face of $M$, and $\phi_i \circ F : R^3 \to R^2 $ is an isometry. Then $M$ is a smooth manifold, we can compute the De Rham cohomology $H^*(\Omega(M;TM),d)$, where $\Omega(M;TM)$ is the set of {\it differential forms}, and $d$ is the {\it exterior derivative}. In our setting, all computations are carried out on meshes, which are piecewise linear. Therefore, it is enough to just use piecewise linear differential forms on $M$. We define the set of piecewise linear forms as

\begin{equation}
\Omega^{PL}(M;TM) = \Omega^{PL}_0 \cup \Omega^{PL}_1 \cup \Omega^{PL}_2
\end{equation}

Here $\Omega^{PL}_0$ is the set of piecewise linear functions on $M$ defined on its vertices,  $\Omega^{PL}_1$ is the set of piecewise constant 1-forms which are consistent along the edges,
\begin{equation}
\label{E:consistence_edge}
\int_{[u,v]} \omega|_{[u,v,w]} = \int_{[u,v]} \omega|_{[t,v,u]}, \omega \in \Omega^{PL}_1,
\end{equation}
where $[u,v,w]$ and $[t,v,u]$ are the two faces adjacent to $[u,v]$, $ \Omega^{PL}_2$ is the set of piecewise constant two forms.

All the computations are defined for De Rham cohomology in concept and for simplicial cohomology in implementation.  We connect differential forms with simplicial cocycles by the following map: Given $\omega\in Z^qK$,
\begin{eqnarray}
\Gamma &:& Z^qK \to \Omega^{PL}_q( M;TM )\\
\omega \sigma &=& \int_\sigma  \Gamma \omega, \forall \sigma\in C_qK.
\end{eqnarray}

It is easy to verify that $\Gamma$ is well defined, one to one and is also commutative to differential operators,
\begin{equation}
d\circ\Gamma = \Gamma \circ \delta.
\end{equation}
So in the following discussion, we do not differentiate simplicial cocycles and piecewise linear differential forms explicitly.

\subsection{Harmonic 1-form and Holomorphic 1-forms}
According to Hodge theory \cite{Yau:lecture}, each cohomology class in $H(\Omega(M;TM),d)$ has a harmonic representative, which minimizes the harmonic energy as defined below. Suppose $f \in \Omega^0(M;TM)$, the harmonic energy of $f$ is
\begin{equation}
\label{E:harmonic_energy_0}
E(f) = \frac{1}{2} \int_M ||df||^2 d\sigma.
\end{equation}
The norm is Euclidean norm, and $d\sigma$ is the area element. The harmonic energy for 1-forms is defined similarly. Suppose $\omega  \in \Omega^1(M;TM)$, the harmonic energy of $\omega$ is
\begin{equation}
\label{E:harmonic_energy_1}
E(\omega) = \frac{1}{2} \int_M ||\omega||^2 d\sigma.
\end{equation}

In the case where $M$ is a mesh, the harmonic energy can be simplified in the format of string energy and defined on $C^*K$. Suppose $f \in C^0K$, the harmonic energy \ref{E:harmonic_energy_0} can be rewritten as
\begin{equation}
\label{E:string_energy_0}
E(f) = \sum_{[u,v]\in K} k_{u,v}|| f(u) - f(v)||^2.
\end{equation}
For 1-form $\omega \in C^1K$, the harmonic energy \ref{E:harmonic_energy_1} is reformulated as
\begin{equation}
\label{E:string_energy_1}
E(\omega) = \sum_{[u,v]\in K} k_{u,v} ||\omega [u,v]||^2.
\end{equation}
Suppose edge $[u,v]$ has two adjacent faces $T_\alpha, T_\beta$, $T_\alpha =
[v_0,v_1,v_2]$, define parameters
\begin{eqnarray}
a_{v_1,v_2}^\alpha &=& \frac{1}{2}\frac{(v_1-v_3)\cdot(v_2-v_3)}{%
(v_1-v_3)\times(v_2-v_3)} \\
a_{v_2,v_3}^\alpha &=& \frac{1}{2}\frac{(v_2-v_1)\cdot(v_3-v_1)}{%
(v_2-v_1)\times(v_3-v_1)} \\
a_{v_3,v_1}^\alpha &=& \frac{1}{2}\frac{(v_3-v_2)\cdot(v_1-v_2)}{%
(v_3-v_2)\times(v_1-v_2)}. \\
\end{eqnarray}
$a_{u,v}^\beta$ can be defined similarly, then
\begin{equation}
\label{E:string_coefficient}
k(u,v) = a_{u,v}^\alpha + a_{u,v}^\beta.
\end{equation}

A function $f \in C^0K$ with local minimum harmonic energy is called a harmonic function. A cocycle $\omega \in C^1K$ with local minimum harmonic energy is called a harmonic form.

The Laplacian operator $\Delta^{PL}:\Omega^{PL}_0\to \Omega^{PL}_0$ is defined as the derivative of $E(f)$ with respect to $f$
\begin{eqnarray}
\label{E:Laplacian}
\Delta^{pl}f |_u &=& \sum_{[u,v]\in K} k_{u,v}( f(u) - f(v) ).
\end{eqnarray}

\subsection{Complex structure}
A 2-dimensional manifold $M$  has a natural complex structure. In our setting where $M$ is a mesh, the complex structure is constructed explicitly in ~\cite{Duchamp:harmonic}.

Any genus zero surface $M$ is conformally equivalent to $S^{2}$. $u:M\rightarrow S^{2}$ is conformal if and only if $u$ is harmonic. The conformal automorphism group of $S^{2}$ is 6 dimensional, which is the
Mobius transformation group defined on the complex plane $\mathbb{C}$. If we fix the images of 3 points, then there is a unique conformal map from $M$ to $S^{2}$.

For non-zero genus surfaces, we study the structure of its holomorphic differential group. The following form
\begin{eqnarray}
\tau + \sqrt{-1}\omega, \tau,\omega \in \Omega^1(M;TM)
\end{eqnarray}
is called a holomorphic form if both $\tau$ and $\sigma$ are harmonic  and $*\tau = \omega$. Here $*$ is the Hodge star operator. Suppose $\{v_1,v_2\}$ are orthonormal bases of a tangent space on $M$, then
\begin{equation}
\omega(v_1) = *\omega(v_2).
\end{equation}

The set of  holomorphic 1-forms is denoted as $H^{1,0}(M,\mathbb{C})$. Let $M$ be a compact Riemann surface of genus $g$ and $B=\{e_1,e_2,\cdots,e_{2g}\}$ be an arbitrary basis of $H_1(M,Z)$. The intersection matrix $C$ of the above basis has entries
\begin{equation}
c_{ij} = - e_i \cdot e_j,
\end{equation}
where the dot denotes the algebraic number of intersections. A basis $B^* =\{\omega_1,\omega_2,\cdots,\omega_{2g} \}$ of the real vector space $H^{1,0}(M,\mathbb{C})$ is the dual of $B$ if
\begin{equation}
Re \int_{e_i} \omega_j = c_{ij}.
\end{equation}
From Riemann bilinear relations ~\cite{Griffiths} it follows that the matrix $S$ with entries
\begin{equation}
Im \int_{e_i} \omega_j = s_{ij}.
\end{equation}
is symmetric and positive definite. The complex structure in $H^{1,0}(M,\mathbb{C})$ is given by a matrix $R$ with respect to the basis $B$ and satisfies $R^2 = -I$. The following relation holds
\begin{equation}
\label{E:RS}
CR = S.
\end{equation}
After Weyl ~\cite{Weyl} and Siegel ~\cite{Siegel}, the matrix $R$ is called the period matrix of $M$ with respect to the basis $B$. Let $a$ be a holomorphic automorphism of $M$, and let $[a]$ denote the matrix of its action on the homology and cohomology with respect to the above basis, then
\begin{equation}
[a]^{-1} R[a] = R, [a]^T C [a] = C.
\end{equation}
The pair $(R,C)$ determines the analytic structure of a given Riemann surface in the following sense: two such pairs, $(R_1,C_1)$ and $(R_2,C_2)$ determine the same structure if and only if there exists an integral matrix $N$ whose determinant is $\pm 1$ such that
\begin{equation}
N^{-1}R_1N = R_2, N^t C_1 N = C_2
\end{equation}
If the bases $B_1$ and $B_2$ are canonical ones, then both $C_1$ and $C_2$ are identities, and $N$ is an integral symplectic matrix.

\section{Conformal mapping for genus zero surfaces}

Given two genus zero meshes $M_1,M_2$, there are many conformal mappings between them. The algorithm for computing conformal mapping is based on the fact that harmonic maps are conformal for genus zero surfaces. All conformal mappings between $M_1,M_2$ form a group, which is the so-called Mobius group. Our method is as follows: first find a homeomorphism $\mathbf{h}$ between $M_1$ and $M_2$, then diffuse $\mathbf{h}$ so that $\mathbf{h}$ minimizes the harmonic energy. In order to ensure the convergence of the algorithm, special constraints are added so that the solution is unique.

\subsection{Constrained Variational Problem}

Suppose $M_1$ and $M_2$ are genus zero meshes, $\mathbf{h} : M_1 \to M_2 $ is a degree one mapping. We would like to minimize the harmonic energy $E(\mathbf{h})$,
\begin{equation}
	E(\mathbf{h}) = \sum_{[u,v]\in K_1} k_{u,v}|| \mathbf{h}(u) - \mathbf{h}(v) ||^2, \mathbf{h} = (h_0,h_1,h_2).
\end{equation}
The Laplacian for $\mathbf{h}$ is simple
\begin{equation}
\Delta^{PL} \mathbf{h} = ( \Delta^{PL}h_0, \Delta^{PL}h_1, \Delta^{PL}h_2 ).
\end{equation}
Then if $\mathbf{h}$ is harmonic, the tangential component of $\Delta^{PL}\mathbf{h}$ is zero. Define projection operator
\begin{equation}
P_{\mathbf{v}} = I - \frac{ \mathbf{v} \otimes \mathbf{v}^T}{ \mathbf{v}^T \mathbf{v} }, \mathbf{v} \in R^3,
\end{equation}
where $\otimes$ is tensor product and $I$ is an identity matrix. Then $\mathbf{h}$ is harmonic if and only if
\begin{equation}
P_{\mathbf{n}\circ\mathbf{h}} \Delta^{PL}\mathbf{h} = 0,
\end{equation}
where $\mathbf{n}$ is the normal on $M_2$.

In order to ensure the process converge to a unique solution, we have to add extra constraints. We force the center of mass of the surface to be at its origin, that is,
\begin{equation}
\label{E:center_mass}
\int_{M_2} \mathbf{h} d\sigma_{M_1} = \mathbf{0},
\end{equation}
where $d\sigma_{M_1}$ is the area element on $M_1$. This constraint will guarantee the solution is unique up to a rotation. Then we can construct the partial differential equation
\begin{equation}
\label{E:spherical_conformal}
\frac{\partial \mathbf{h} }{\partial t} + P_{\mathbf{n}\circ \mathbf{h}}  \Delta^{PL} \mathbf{h}  = 0
\end{equation}
with constraints \ref{E:center_mass}. The steady state solution of $\mathbf{h}$ is the conformal mapping from $M_1$ to $M_2$. Equation \ref{E:spherical_conformal} can be solved by iterative methods.

\subsection{ Steepest Descendent Algorithm}
In our implementation, we fix $M_2$ as $S^2$. In order to compute the initial homeomorphism from $M_1$ to $S^2$, we first compute the spherical barricentric embedding, which minimizes the barricentric string energy. The barricentric energy is defined as in \ref{E:string_energy_0}, where we let
\begin{equation}
k_{u,v} \equiv 1.
\end{equation}
The corresponding Laplacian is defined as \ref{E:Laplacian} with constant unit $k_{u,v}$. Then the following algorithm computes spherical barricentric embedding,

\noindent \texttt{\\
\indent {\bf Input} mesh $M$, step length $\delta t$, threshold $\epsilon$.\\
\indent {\bf Output} sphereial barricentric mapping $\mathbf{h}$.\\
\\
\indent 1. Compute Gauss map $\mathbf{n}$ from $M$ to $S^2$,$\mathbf{h}\leftarrow \mathbf{n}$. \\
\indent 2. Compute barricentric energy $E(\mathbf{h})$, if $\delta E < \epsilon$ then return $\mathbf{h}$.\\
\indent 3. Compute tangential Laplacian of $\mathbf{h}$, $\delta h \leftarrow P_{\mathbf{n}\circ \mathbf{h}}\Delta^{PL}\mathbf{h}$\\
\indent 4. Update $\mathbf{h}$ by $\mathbf{h} \leftarrow \mathbf{h} - \delta t \times \delta \mathbf{h}$.\\
\indent 5. Repeat 2 through 4.\\
}

\begin{center}
Algorithm 1. Spherical barricentric embedding
\end{center}
\medskip

In practice, barricentric embedding converges faster than spherical harmonic embedding, and there are no extra constraints. Hence we use it as the initial embedding to compute spherical conformal mapping. The spherical conformal embedding algorithm is more complicated. In each iteration an extra normalization step is inserted so that the mass center of the surface stays in the origin during the whole process.

\noindent \texttt{\\
\indent {\bf Gu-Yau Algorithm } for genus zero mesh.\\
\indent {\bf Input}  mesh $M$, step length $\delta t$, threshold $\epsilon$.\\
\indent {\bf Output} spherical conformal map $\mathbf{h}$.\\
\\
\indent 1. Compute spherical barricentric map, $\mathbf{b}$ from $M$ to $S^2$, $\mathbf{h}\leftarrow b$. \\
\indent 2. Compute harmonic energy $E(\mathbf{h})$, if $\delta E < \epsilon$  then return $\mathbf{h}$.\\
\indent 3. Compute tangential Laplacian of $\mathbf{h}$, $\delta \mathbf{h} \leftarrow P_{\mathbf{n}\circ \mathbf{h}} \Delta^{PL}\mathbf{h}$.\\
\indent 4. Update $\mathbf{h}$ by $\mathbf{h} \leftarrow \mathbf{h} - \delta t \times \delta \mathbf{h}$.\\
\indent 5. Compute a Mobius transformation $\mathbf{m}$, such that $\mathbf{m}\circ \mathbf{h}$ satisfies the center \\
\indent \indent \indent of mass constraint equation \ref{E:center_mass}.\\
\indent 6. Repeat 2 through 5.\\
}
\begin{center}
Algorithm 2. Spherical Conformal Embedding
\end{center}
\medskip

In step 5 above, the Mobius transformation on $S^2$ is in the form $\phi^{-1}\circ f \circ \phi$, where $\phi$ is the stereo-graphic projection from $S^2$ to the complex plane.
\begin{equation}
\phi(x_0,x_1,x_2) = (\frac{x_0}{1+x_2},\frac{x_1}{1+x_2}), (x_0,x_1,x_2) \in R^3\\
\end{equation}
$f$ is a Mobius transformation on $\mathbb{C}$,
\begin{equation}
f(z) = \frac{az+b}{cz+d}, a,b,c,d \in \mathbb{C}, ad - bc \neq 0
\end{equation}
In practice, it is expensive to normalize $\mathbf{h}$ by Mobius transformation, we simply shift the center of mass of $\mathbf{h}(M_1)$ to the origin and normalize $\mathbf{h}(v), v\in K$ to the unit vector.

Figure (a) shows a conformal mapping from a bunny model to a sphere, and the bunny is texture mapped using the spherical coordinates as texture parameters.

\section{Computing Conformal Structure for non-zero genus meshes}
\subsection{Overview}
For non-zero genus meshes, the computation of conformal structure is much more complicated. The goal is to find the complete bases of the holomorphic 1-form group. The algorithm can be summarized in the following steps:

\noindent \texttt{\\
\indent {\bf Gu-Yau Algorithm} for non-zero genus mesh\\
\indent {\bf Input} a mesh $M$.
\indent {\bf Output} a set of bases of holomorphic differentials.\\
\\
\indent 1. Compute homology group bases $B=\{e_1,e_2,\cdots,e_g,e_g+1\cdots, e_{2g}\}$. \\
\indent 2. Compute cohomology group bases $\Omega=\{\omega_1,\omega_2,\cdots, \omega_{2g}\}$, which are the dual \\
\indent \indent \indent of $B$. \\
\indent 3. Compute harmonic 1-forms $\zeta=\{\zeta_1,\cdots,\zeta_{2g}\}$, such that $\zeta_i$ is homologous to $\omega_i$. \\
\indent 4. Apply hodge star on $\zeta_i$, and compute holomorphic 1-forms $\zeta_i + \sqrt{-1}(^*\zeta_i)$.\\
}
\begin{center}
Algorithm 3. Compute Holomorphic Differentials
\end{center}

The following subsections explain each step in details.
\subsection{Computing Homology }
There are many methods for computing homology groups $H_*K$ of a simplical complex $K$. In our implementation, we use the classic algorithm, which is based on reducing boundary operator matrices $\partial_q$ to their Smith normal form ~\cite{Mun84}. In order to avoid the substantial computational cost of the reduction to Simth normal form, the mesh is simplified by using progressive mesh algorithm introduced in ~\cite{HH:PM}. Once the homology bases $B$ are found on the coarser mesh, they are mapped back to the finer mesh through a sequence of vertex splits. At each vertex split step, we check the neighborhood of current split vertex, and preserve the connectness of each homology base cycle in $B$. Finally, on the finer mesh, we use Dijkstra algorithm to shorten each base cycle, and perturb them such that they intersect transversely.

The fundamental domain is also computed by the retraction algorithm described in \cite{Gu:Geometry}. The following is the basic procedure: at the beginning, we remove one aritrary face, record the boundary. At each step we remove one face attached to the current boundary, all the removed faces always form a topological disk. The boundary of this disk is kept and updated until all faces are removed. Then we cut the mesh along the final boundary to get the fundamental domain.

\subsection{Computing Cohomology}

Once we obtain homology bases set $B$, we can compute the cohomology bases set $\Omega$ dual to $B$, such that
\begin{equation}
\int_{e_i} \omega_j = \delta_{ij}
\end{equation}

We chose a handle and the pair of conjugate homology cycles on it, denoted as $\{e_i,e_{i+g}\}$. Then we split the mesh along these 2 cycles. Next, we map the boundary to the boundary of a unit square, and map the interior of the mesh to the unit square by Floater embedding algorithm as described in ~\cite{Floater}. Then the 1-forms $\{dx,dy\}$ are the duals of $\{e_i,e_{i+g}\}$.

\noindent \texttt{\\
\indent {\bf Input} mesh $M$, a pair of cycles $\{a,b\}$, such that $a\cap b = 1$. \\
\indent {\bf Output} $\{\omega_a, \omega_b\} \in C^1K$, dual of  $\{a,b\}$.\\
\\
\indent 1. Slice mesh $M$ open along  $\{a,b\}$. $\partial M = aba^{-1}b^{-1}$. \\
\indent 2. Map $aba^{-1}b^{-1}$ to the boundary of $D=[0,1]\times [0,1]$. \\
\indent 3. Map interior of $M$ to $D$ by Floater embedding.\\
\indent 4. Return $\omega_a \leftarrow dx$, $\omega_b \leftarrow dy$.\\
}

\begin{center}
Algorithm 4. Compute Cohomology
\end{center}

\subsection{Computing Harmonic Forms}
Suppose the cohomology bases of mesh $M$ are $\Omega=\{ \omega_1,\omega_2, \cdots, \omega_{2g} \}$, we deform them to harmonic forms by adding exact 1-forms $\delta f_i$, where $f_i \in C^0K$, such that $\omega_i + df_i$ minimizes the harmonic norm in equation \ref{E:string_energy_1}.

\noindent \texttt{\\
\indent {\bf Input} 1-form $\omega \in C^1K$. \\
\indent {\bf Output} harmonic 1-form $\omega$.\\
\\
\indent  1. $F \leftarrow 0$. \\
\indent  2. Compute Laplacian
\begin{equation}
\Delta^{PL} F = \sum_{[u,v]\in K} k_{u,v} ( F(u) - F(v) + \omega [u,v])
\end{equation}\\
\indent 3. $F \leftarrow F - \Delta F \times \delta t$. \\
\indent 4. Compute harmonic energy $E(\omega+\delta F)$, if $\delta E < \epsilon$ then $\omega \leftarrow \omega + \delta F$, return. \\
\indent 5. Repeat 2 through 4. \\
}
\begin{center}
Algorithm 5. Compute Harmonic Forms
\end{center}
This is the most time-consuming step during the whole procedure. In practice, we perform local optimization.

\subsection{Computing Holomorphic Forms}
Given a set of harmonic 1-form bases $\Omega=\{\omega_1,\cdots,\omega_{2g}\}$, we can construct the bases of  holomorphic 1-forms directly by pairing $\omega_i$ with its Hodge star ${\,}^*\omega_i$.
Given $\omega \in C^1K$, then $\Gamma \omega \in \Omega^{PL}_1(M;TM)$, ${\,}^* \Gamma \omega$ is formulated by:

\begin{eqnarray}
	\Gamma \omega  &=& f dx + g dy \\
	\label{E:star} {\,}^*\Gamma \omega  &=& f dy - g dx
\end{eqnarray}
Here $(x,y)$ are local coordinates as defined in equation \ref{E:chart}, $f,g$ are constants on each face of $M$.

Hodge star transforms harmonic forms to harmonic forms. If $\omega$ is harmonic, then ${\,}^*\omega$ is also harmonic, and it can be represented as a linear combination of $\omega_i$'s. Suppose
\begin{equation}
{\,}^*\omega = \sum_{i=1}^{2g} \alpha_i \omega_i,
\end{equation}
Then we can compute the integration of wedge product
\begin{equation}
\label{E:wedge_relation}
\int_M \omega_i \wedge {\,}^* \omega = \int_M \Gamma \omega_i \wedge {\,}^* (\Gamma \omega), ~i = 1, 2, \cdots, 2g
\end{equation}

Equation \ref{E:wedge_relation} can be formulated as the following linear system
\begin{equation}
\label{E:linear}
\mathbf{A}\mathbf{\alpha} = \mathbf{b},
\end{equation}
where $\mathbf{\alpha}=(\alpha_i)$, matrix $\mathbf{A}$ is with entries
\begin{equation}
a_{ij} = \int_M \omega_i \wedge \omega_j.
\end{equation}
Because $\omega_i$ are dual cocyles of $e_i$, so
\begin{equation}
a_{ij} = e_i \cap e_j.
\end{equation}
Vector $\mathbf{b}$ has entries
\begin{equation}
b_i = \int_M \Gamma \omega_i \wedge {\,}^*(\Gamma \omega).
\end{equation}
Assume $\Gamma \omega_i=fdx + gdy$ , $\Gamma \omega = pdx + qdy$, from \ref{E:star}
\begin{equation}
b_i = \sum_{[u,v,w]\in K}(fp+gq)\sigma_{[u,v,w]}.
\end{equation}
By our construction, matrix $A$ is also the intersection matrix of homology bases $R$, so $A$ is non-degenerated. ${\,}^*\omega$ is uniquely determined. The following is the algorithm to compute holomorphic 1-forms:

\noindent\texttt{\\
\indent {\bf Input} Bases of harmonic 1-form group, $\{\omega_1,\omega_2,\cdots,\omega_{2g}\}$,a harmonic 1-form $\omega$.\\
\indent {\bf Output} Holomorphic 1-form $\omega + \sqrt{-1} {\,}^* \omega$.\\
\\
\indent 1. Compute $\Gamma \omega$ and $\Gamma \omega_i$.\\
\indent 2. Compute ${\,}^* \Gamma \omega$.\\
\indent 3. Compute $b_i$. \\
\indent 4. Solve linear system \ref{E:linear}.\\
\indent 5. Return $\zeta = \omega + \sqrt{-1} {\,}^*\omega$.
}
\begin{center}
Algorithm 6. Compute Holomorphic Forms
\end{center}

By applying the above algorithm, we can compute the bases of holomorphic differentials of $M$. Suppose we treat the holomorphic differentials as a complex vector space, we denote a set of bases  as $\{\zeta_1,\zeta_2,\cdots,\zeta_g\}$, where $g$ is the genus of $M$.
The figure shows the results of computing holomorphic 1-forms on meshes. Figure (b) shows the result for a genus one mesh. By integrating a holomorphic 1-form, the mesh is mapped to the plane. Then a checker board is texture mapped to the mesh using the plane as the texture parameter space. Figure (c) is constructed similarly for a genus two surface.

By linearly combining $\zeta_i$s, we can construct all holomorphic 1-forms on $M$. By integrating holomorphic 1-forms on the fundamental domain, the mesh is globally conformally mapped to the plane with finite singularities. The number of singularities on $M$ is $2g-2$.

\section{Performance Analysis}

The algorithm is independent of the choices of geometric realization of homology cycles, but dependent on their homology classes. In a future paper, we will give a method to compute global conformal parametrization which is independent of the choice of homology classes too. 

It is obvious that the extruding parts, like the ears of the bunny are mapped to relatively small regions. Those planar regions are very dense. During the optimization process, these regions converge more slowly. In general, special local optimization is necessary for these regions.

The energy form $k_{u,v}||f(u)-f(v)||^2$ is determined by $k_{u,v}$. During our experiments, we find that if $k_{u,v}$ are all positive, then the algorithm converges faster. For the harmonic energy minimization, the edge coefficients \ref{E:string_coefficient} can be reformulated as
\begin{equation}
k_{u,v}=cot \angle\alpha + cot \angle\beta.
\end{equation}
Here there are two faces sharing edge $[u,v]$ and $\alpha,\beta$ are the two angles in these faces opposite to the edge. In our implementation, we carry out some preprocessing on meshes, to swap or split edges with negative $k_{u,v}$. This process improves the convergence speed.

\section{Applications}

\subsection{Computational Topology}
Homology has a group structure, cohomology has a ring structure, so cohomology can convey more geometric information of the manifolds. The cohomology bases can be used to detect the homology class of a closed curve. Suppose a set of cohomology bases $\{\omega_1,\omega_2,\cdots,\omega_{g},\omega_{g+1},\cdots,\omega_{2g}\}$ has been computed, given an arbitrary closed curve $r$, if $r$ is homologous to zero, then the following must hold
\begin{equation}
\int_r \omega_i = 0, \forall i 
\end{equation}
Once $r$ is homologous to zero, we can find the domain whose boundry is $r$ by the following simple flooding algorithm. First we label all the faces on the left of $r$ and adjacent to $r$. Then we label all the neighboring faces to them. We repeat this process, until no further face can be labelled. Then all the labelled faces form the domain.

\subsection{Geometry Matching}
Conformal structure is determined by Riemann metric, so it is independent of triangulation. Conformal structure is stable in the sense that if we perturbate the metric, the conformal structure changes continuously. Therefore, it is tolerant of noises and not sensitive to different resolutions.  We perform some numerical experiments to verify this property of conformal mappings. Suppose we have 2 geometrical similar surfaces $M_1,M_2$, in order to find the best geometric match, we can conformally map them to a canonical domain $D$,

\begin{equation}
\begin{diagram}
\node{M_1} \arrow[2]{e,t}{f_2^{-1}\circ f_1} \arrow{se,r}{f_1} \node[2]{M_2} \arrow{sw,r}{f_2}\\
\node[2]{D}
\end{diagram}
\end{equation}

Then $f_2 \circ f_1$ gives the desired geometric matching. In this process, the appropriate boundary conditions should be set up correctly. 

\subsection{Geometry Classification}

The non-zero genus surfaces can be classified by their conformal structures naturally. After the bases of holomorphic 1-form group are computed, it is straightforward to compute the period matrices. During the construction of homology bases, we can obtain a canonical set of homology bases, that is
\begin{equation}
\left\{
\begin{array}{lll}
r_i \cap r_{g+i} &=& +1, i = 1, 2, \cdots, g \\
r_i \cap r_j     &=&  0, j \neq {g+i}
\end{array}
\right.
\end{equation}
Then the period matrix is
\begin{equation}
\mathbf{P}=
\begin{pmatrix}
\int_{r_1} \zeta_1 & \int_{r_1} \zeta_2 & \cdots & \int_{r_1} \zeta_{2g-1}&\int_{r_1} \zeta_{2g} \\
\int_{r_2} \zeta_1 & \int_{r_2} \zeta_2 & \cdots & \int_{r_2} \zeta_{2g-1}& \int_{r_2} \zeta_{2g} \\
\hdotsfor{5}\\
\int_{r_{2g}} \zeta_1 & \int_{r_1} \zeta_2 & \cdots & \int_{r_{2g}} \zeta_{2g-1}& \int_{r_{2g}} \zeta_{2g} \\
\end{pmatrix}
\end{equation}
If two surfaces $M_1,M_2$ are conformally equivalent, then there exists an integral symplectic matrix $N$, such that $N^{-1}P_1 N = P_2$. $N$ is the homology bases transformation matrix. 

\subsection{ Global conformal parametrization}
A mesh can be parameterized conformally by integrating holomorphic 1-forms on it. The parametrization is globally conformal except for finite singularities. By changing holomorphic 1-forms, the neighborhoods of singularities can be conformally parametrized too.

By using conformal parameters, many important geometric quantities which are valuable for geometric analysis can be computed explicitly. 

\section{Conclusion}
This paper introduces a systematic way to compute conformal structure for general surfaces represented as triangle meshes. The homology is computed by simplicial complex structure. The dual cohomology bases are constructed explicitly. Each cohomology cocyle is diffused to a harmonic 1-form by adding an exact 1-form to minimize the harmonic energy. The Hodge star operation is carried out on the harmonic forms by solving a linear system. Then the bases of holomorphic differentials are constructed. To the best of our knowledge, this paper is the first one to solve this problem completely. The methods introduced here are very general. The harmonic 1-forms, holomorphic 1-forms have much broader applications. The conformal structure can be applied in many theoretic fields as well as engineering fields.

\section{Future Research}

Conformal structures of closed surfaces are studied thoroughly in this paper. We would like to generalize the results to open surfaces. Current computations are based on mesh structures. We will generalize the algorithms to other surface representations, such as implicit surfaces and level sets. The optimization of harmonic energy is computationally expensive. In the future, we will use multi-resolution methods to improve the speed. We will explore more on the relation between the eigenvalues, eigenfunctions of Laplacian operator and geometry. Current conformal parametrization is dependent on the choices of homology bases. In a future paper, we will introduce a new method which is independent of those choices.

\bibliographystyle{acm}

\end{document}